\documentclass[floatfix,aps,prl,preprint,superscriptaddress]{revtex4}
\usepackage[]{epsfig}
\usepackage{times,amsmath,amssymb}

\usepackage{graphicx,color}
\usepackage{hyperref}
\hypersetup{
colorlinks=true, 
urlcolor=blue
}

\begin{document}

\title{Visual explanations of machine learning model estimating charge states in quantum dots}

\author{Yui Muto}
\affiliation{Research Institute of Electrical Communication, Tohoku University, 2-1-1 Katahira, Aoba-ku, Sendai 980-8577, Japan}
\affiliation{Department of Electronic Engineering, Graduate School of Engineering, Tohoku University, Aoba 6-6-05, Aramaki, Aoba-Ku, Sendai 980-8579, Japan}

\author{Takumi Nakaso}
\affiliation{LeapMind, 28-1 Maruyama-cho, Shibuya-ku, Tokyo 150-0044, Japan}

\author{Motoya Shinozaki}
\affiliation{WPI Advanced Institute for Materials Research, Tohoku University, 2-1-1 Katahira, Aoba-ku, Sendai 980-8577, Japan}

\author{Takumi Aizawa}
\affiliation{Research Institute of Electrical Communication, Tohoku University, 2-1-1 Katahira, Aoba-ku, Sendai 980-8577, Japan}
\affiliation{Department of Electronic Engineering, Graduate School of Engineering, Tohoku University, Aoba 6-6-05, Aramaki, Aoba-Ku, Sendai 980-8579, Japan}

\author{Takahito Kitada}
\affiliation{Research Institute of Electrical Communication, Tohoku University, 2-1-1 Katahira, Aoba-ku, Sendai 980-8577, Japan}
\affiliation{Department of Electronic Engineering, Graduate School of Engineering, Tohoku University, Aoba 6-6-05, Aramaki, Aoba-Ku, Sendai 980-8579, Japan}

\author{Takashi Nakajima}
\affiliation{Center for Emergent Matter Science, RIKEN, 2-1 Hirosawa, Wako, Saitama 351-0198, Japan}

\author{Matthieu R. Delbecq}
\affiliation{Center for Emergent Matter Science, RIKEN, 2-1 Hirosawa, Wako, Saitama 351-0198, Japan}

\author{Jun Yoneda}
\affiliation{Center for Emergent Matter Science, RIKEN, 2-1 Hirosawa, Wako, Saitama 351-0198, Japan}

\author{Kenta Takeda}
\affiliation{Center for Emergent Matter Science, RIKEN, 2-1 Hirosawa, Wako, Saitama 351-0198, Japan}

\author{Akito Noiri}
\affiliation{Center for Emergent Matter Science, RIKEN, 2-1 Hirosawa, Wako, Saitama 351-0198, Japan}

\author{Arne Ludwig}
\affiliation{Ruhr University, Bochum, Universitätsstraße 150, 44801 Bochum, German}

\author{Andreas D. Wieck}
\affiliation{Ruhr University, Bochum, Universitätsstraße 150, 44801 Bochum, German}

\author{Seigo Tarucha}
\affiliation{Center for Emergent Matter Science, RIKEN, 2-1 Hirosawa, Wako, Saitama 351-0198, Japan}

\author{Atsunori Kanemura}
\affiliation{LeapMind, 28-1 Maruyama-cho, Shibuya-ku, Tokyo 150-0044, Japan}

\author{Motoki Shiga}
\affiliation{Unprecedented-scale Data Analytics Center, Tohoku University, 6-3 Aoba, Aramakiaza, Aoba-ku, Sendai, 980-8578 Japan}
\affiliation{RIKEN Center for Advanced Intelligence Project, 1-4-1 Nihonbashi, Chuo-ku, Tokyo 103-0027, Japan}
\affiliation{Graduate School of Information Science, Tohoku University, 6-3-09 Aoba, Aramaki-aza Aoba-ku, Sendai, 980-8579, Japan}

\author{Tomohiro Otsuka}
\email[]{tomohiro.otsuka@tohoku.ac.jp}
\affiliation{WPI Advanced Institute for Materials Research, Tohoku University, 2-1-1 Katahira, Aoba-ku, Sendai 980-8577, Japan}
\affiliation{Research Institute of Electrical Communication, Tohoku University, 2-1-1 Katahira, Aoba-ku, Sendai 980-8577, Japan}
\affiliation{Department of Electronic Engineering, Graduate School of Engineering, Tohoku University, Aoba 6-6-05, Aramaki, Aoba-Ku, Sendai 980-8579, Japan}
\affiliation{Center for Science and Innovation in Spintronics, Tohoku University, 2-1-1 Katahira, Aoba-ku, Sendai 980-8577, Japan}
\affiliation{Center for Emergent Matter Science, RIKEN, 2-1 Hirosawa, Wako, Saitama 351-0198, Japan}

\date{\today}

\begin{abstract}
Charge state recognition in quantum dot devices is important in the preparation of quantum bits for quantum information processing. Toward auto-tuning of larger-scale quantum devices, automatic charge state recognition by machine learning has been demonstrated. For further development of this technology, an understanding of the operation of the machine learning model, which is usually a black box, will be useful. In this study, we analyze the explainability of the machine learning model estimating charge states in quantum dots by gradient-weighted class activation mapping, which identified class-discriminative regions for the predictions. The model predicts the state based on the change transition lines, indicating that human-like recognition is realized. We also demonstrate improvements of the model by utilizing feedback from the mapping results. Due to the simplicity of our simulation and pre-processing methods, our approach offers scalability without significant additional simulation costs, demonstrating its suitability for future quantum dot system expansions.
\end{abstract}

\maketitle

\section{Introduction}

Semiconductor quantum bits (qubits) that utilize electron spins in semiconductor quantum dots are expected to be good candidates for future qubits because of their high operation fidelity and excellent integration properties. 
Basic operations such as single-qubit\cite{Koppens2006, PioroLadriele2008} and two-qubit operations\cite{Veldhorst2015, Watson2018} have been demonstrated, and quantum error correction has also been demonstrated\cite{Takeda2022} with improved operation fidelity\cite{Yoneda2014, Veldhorst2014, Takeda2016, Yoneda2018, Noiri2022, Madzik2022, Xue2022}. 
Furthermore, attempts are being made to construct large-scale quantum systems using semiconductor integration technology\cite{Maurand2016, Vandersypen2017, Veldhorst2017, Camenzind2022, Zwerver2022, Otsuka2016, Ito2016, Volk2019}.
It is essential to trap one electron in each quantum dot to construct semiconductor qubits. 
Charge state tuning in semiconductor quantum dot devices is required for this purpose. 
In previous experiments, this has been mainly accomplished by manually adjusting the gate voltages to achieve the desired charge states. 
However, this approach requires a significant amount of time to learn and execute. 
This will make tuning large-scale semiconductor quantum systems difficult in the future.

There is a movement to solve such a difficult problem automatically\cite{Hu2023}.  
The main types of algorithms studied in these methods are script-based algorithms\cite{Baart2016, Diepen2018, Mills2019, Lapointe-Major2020} or machine-learning (ML) methods\cite{Kalantre2019, Teske2019, Lennon2019, Zwolak2020, Darulova2020, Durrer2020, Moon2020, Esbroeck2020, Mei2021, Nguyen2021, Matsumoto2021, Ziegler2022}. 
In particular, ML methods are expected to be more easily applicable to different experimental environments and more compatible with different devices.
Efforts to develop ML methods for auto-tuning quantum device parameters are currently underway~\cite{Zwolak2021, Liu2022}.
In ML methods, the amount of labeled data required for network training is enormous, making the preparation of labeled experimental data as training data a labor-intensive task. 
Consequently, simulations have been utilized to generate training data. 
This approach has led to the development of a method that utilizes the Thomas-Fermi model and deep convolutional neural networks (CNN) for recognizing charge states~\cite {Kalantre2019}. 
Also, by adding various types of noise that could be observed in real experiments to the training data, the noise-robust charge states recognition has been demonstrated~\cite{Ziegler2022}.

Thus far, charge state recognition has primarily been demonstrated in double quantum dots~\cite{Kalantre2019, Zwolak2020, Ziegler2022}. 
Looking towards the future large-scale integration of quantum dots, there is a need to explore scalable methods. 
Also, while the Thomas-Fermi model with added noise is highly useful as a physical model, the necessity of such physical accuracy in a data generation model for machine learning has not been extensively debated. 
For instance, humans tend to estimate charge states by focusing mainly on charge transition lines, without giving much consideration to background noise. 
Therefore, generating training data with a physical model that reflects these characteristics could lead to the realization of an estimator with judgment criteria similar to those of humans. 
Such an estimator would likely be more user-friendly for humans, as it would share the same criteria for judgment. 
To achieve this, it is important to clarify the criteria for judgment, which is usually a black box.

In this study, we analyze the operation of the ML model that estimates the charge states in double quantum dots. 
We propose pre-processing the data through binarization, which enables us to use a simple simulation model to generate training data.
We also visualize the interest of the model by gradient-weighted class activation mapping (Grad-CAM)~\cite{Selvaraju2020} and confirm that they are on the charge transition lines.
By incorporating feedback from the Grad-CAM results, we demonstrate improvements in the ML model.

\section{Preparation of the charge state estimator}

\begin{figure*}
\begin{center}
  \includegraphics{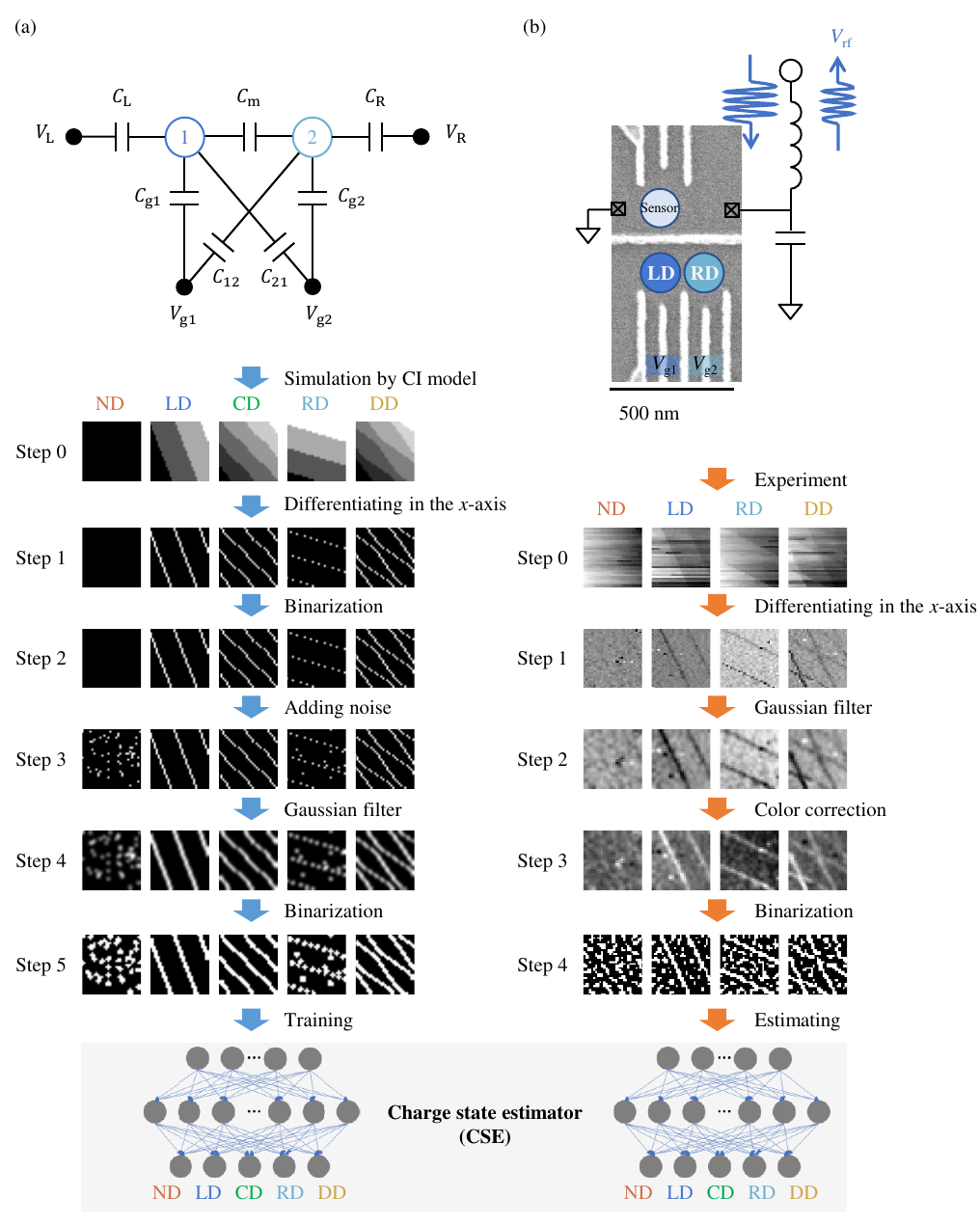}
  \caption{(a) The flow of generating the CSE. 
  Training data for the CNN is prepared by simulation using the CI model. We simplify the data with five pre-processing and then train the CNN. 
  (b) The charge recognition flow of the experimental data measured by radio-frequency reflectometry.
  Steps 1, 2, and 4 are common processes shared with the training data.
  We add the process that changes the color of all charge transition lines into white by utilizing the skewness (step 3).
  }
  \label{fig1}
\end{center}
\end{figure*}

Figure~\ref{fig1}(a) illustrates the process flow for generating a charge state estimator (CSE). 
Initially, the preparation of training data is needed to train the ML model.
We utilize a constant interaction (CI) model in the simulation for simplicity, where quantum dots are modeled as a capacitor circuit\cite{2001KouwenhovenRPP, 2003ElzermanPRB}.
Using this CI model and treating the electrostatic coupling between the double quantum dot and the charge sensor, we can obtain the charge stability diagrams with the size ($30\times30$) pixel measured using a charge sensor by changing the parameters in the model. 
This size is chosen to achieve a sufficient resolution to distinguish the charge transition lines. The next step is to prepare the charge state label: no quantum dot (ND), left (LD),  center (CD), right (RD), or double quantum dot (DD), for each stability diagram (Fig.~\ref{fig1}(a) step 0). 
Note that the CD state is formed by strong coupling between both dot sites, which in the CI model is expressed by setting a large value for $C_{\mathrm m}$.
Here, the diagrams are automatically categorized based on the largest number of quantum dots in the region. 

Next, pre-processing is performed on these images. 
All images are differentiated in the $x$-axis to make the charge transition lines clear (Fig.~\ref{fig1}(a) step 1). 
Then, the images are binarized by adaptive thresholding to simplify them into black and white images (Fig.~\ref{fig1}(a) step 2). 
Here, the charge transition line is set to white, while the background is set to black. 
In this process, we use Adaptive Thresholding by the OpenCV module~\cite{OpenCV}.

To adapt to noisy experimental data, random noise is added to these images. 
Due to binarization in the previous step, noise can be realized by simply flipping the black and white pixel values at random locations (Fig.~\ref{fig1}(a) step 3). 
We adjust the amount of noise such that the number of white pixels after this process does not depend on the charge states. 
Images labeled DD will have more white points than others without this adjustment because there are many charge transition lines, and the CNN will be trained to estimate the state by counting the number of white pixels. We adjust the number of white points in an image between 30 and 110, combining the noise and the charge transition lines. Next, a $3\times3$ Gaussian filter is applied to the images (Fig.~\ref{fig1}(a) step 4). 
The same filter is applied to the experimental data in the pre-processing step. 
After applying the Gaussian filter, the image is again binarized by adaptive thresholding, as shown in Fig.~\ref{fig1}(a) step 5.

Here, 84,325 images, each measuring ($30\times30$) pixel, are undersampled from a total of 180,000 images in the simulation and trained by CNN.
Images for each charge state are prepared in the same quantity.
The details of the CNN are presented in the supplementary information.

\section{Estimation of the charge states}

\subsection{Pre-processing of experimental data}

In this section, we perform the charge state recognition of the real experimental data with the obtained CSE. 
The experimental data is a radio-frequency reflectometry-measured charge sensing signal from a GaAs double quantum dot\cite{2007ReillyAPL, 2021Shinozaki}. 
The size of the experimental data is ($30\times30$) pixel (Fig.~\ref{fig1}(b) step 0). 
Note that the gate voltage range is adjusted to include several charge transition lines in the images.
This prevents the situation where the ($30\times30$) pixel is small compared to the interval of the charge transition lines, which could lead to erroneous recognition of the ND state for other charge states.

Prior to inputting the experimental data into the CSE, it undergoes pre-processing using almost the same methods as the training data. 
This includes differentiation, application of a Gaussian filter, and binarization. 
These steps correspond to steps 1, 2, and 4 in Fig~\ref{fig1}(b), respectively.
Adaptive thresholding can binarize even with a non-uniform background because the threshold is calculated for each small region. 
It is useful for the actual experiments, since charge stability diagrams often have a nonuniform background due to the shift of the charge sensor condition. 
Here, the adaptive thresholding parameters are set to the same as those in the training process.

As an additional processing step (Fig.~\ref{fig1}(b) step 3) for the experimental data, a procedure is applied to ensure that the color of the charge transition lines (black or white) matches that of the simulation data.
The charge transition lines in experiments might be black as shown in Figs.~\ref{fig1}(b) steps 1 and 2, depending on the condition of the charge sensor. 
In this case, we need to invert the color because the lines are white in the training data. 
To detect this case, we utilize the skewness of the image. The skewness $S_{\mathrm{k}}$ is obtained from the histogram of the pixel values by the following equation, 
\begin{equation}
S_{\mathrm{k}} = \frac{\frac{1}{n}\sum_{i=1}^n (x_{\mathrm{i}}-\overline{x})^3}
                      {s^3},
\end{equation}
where $n$ is the number of data points ($30\times30$) , $x_{\mathrm{i}}$ the value of each pixel, $\overline{x}$ the mean of the value, and $s$ the standard deviation. 
Because the number of pixels for the charge transition lines is less than that of the background in the stability diagram, the color of the charge transition lines is reflected in the sign of the skewness. In the case of the black charge transition lines, the skewness becomes negative. Then, we inverted the pixel values. In the calculation of the skewness, the pixel histograms are flattened using CLAHE (Contrast Limited Adaptive Histogram Equalization) to improve noise immunity. This inversion process turns the charge transition lines white depending on the skewness (Fig.~\ref{fig1}(b) step 3). 

\subsection{Charge state recognition}

\begin{figure}
\begin{center}
  \includegraphics{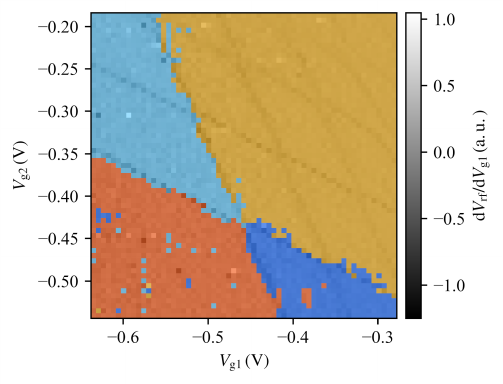}
  \caption{The result of the charge state recognition of experimental data. The color represents the classification result. The orange dots indicate points estimated as ND, the deep blue, green, light blue, and yellow dots indicate LD, CD, RD, and DD, respectively.
  }
  \label{fig2}
\end{center}
\end{figure}

We now discuss the performance of the charge state recognition using the obtained CSE on the experimental data pre-processed by the steps in Fig.~\ref{fig1}(b). 
The experimental data (step 1 in pre-process) is presented in the background layer of Figure~\ref{fig2} using gray-scale representation.
The color in Fig.~\ref{fig2} represents the result of the charge state recognition by the CSE. 
Here, each recognition is performed by extracting each ($30\times30$) pixel area from the original image. 
Subsequently, CSE loads each area and returns the output as the maximum number of quantum dots in the area.  
Orange, deep blue, green, light blue, and yellow dots indicate the points classified as the ND, LD, CD, RD, and DD regions, respectively.
Note that the CD state is not obtained in the used experimental data.
By comparing the results with experimentalists’ classification per pixel, the accuracy is evaluated as $(93.3 \pm 0.8)\%$ with 30 CSEs using the same network architecture.
This means that the model even trained by the simple binarized data can recognize the charge states of noisy experimental data.
We also apply the CSE to the simulation data with noise, and an accuracy of $(97.0 \pm 0.2)\%$ is achieved.
Therefore, the experimental data is estimated with a degree of accuracy comparable to that of the simulation data.
To verify that the CSE developed in this study is not exclusively optimized for only our experimental results, we apply it to publicly available experimental data including CD state from previous studies~\cite{NIST_dataset}.
The CSE demonstrates a $(86.0 \pm 1.2)\%$ accuracy in recognizing charge states, even when applied to data from other group.

\subsection{Visualization of the machine learning model and its feedback}

\begin{figure}
\begin{center}
  \includegraphics{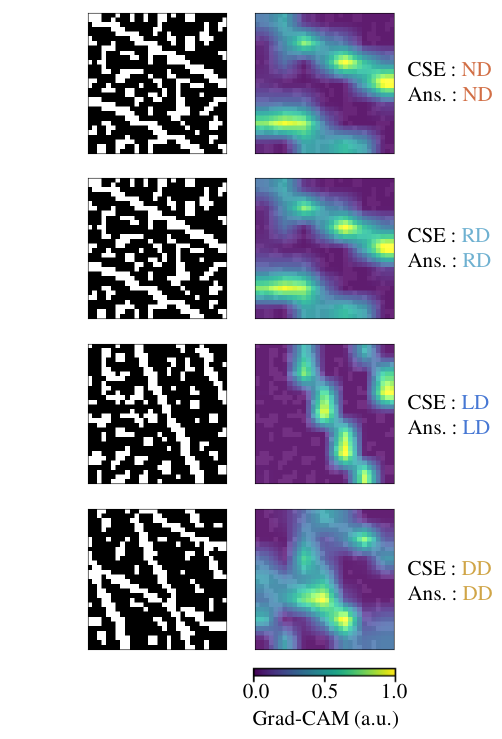}
  \caption{The results of Grad-CAM for real experimental data with estimation result by the CSE.}
  \label{fig3}
\end{center}
\end{figure}

We analyze the explainability of the operation of the CSE on the experimental data using Grad-CAM~\cite{Selvaraju2020}. 
This method identifies class-discriminative regions for each estimation by the model from an input image. 
The results of the Grad-CAM analysis are demonstrated in Fig.~\ref{fig3}. 
The images in the left column are the input images, while those in the right are the outputs from Grad-CAM. 
The brighter regions in the right figures correspond to the regions that affect the result of the CSE. 
We can see bright regions around the charge transition lines in these images.
In the case of the simulation data with noise, the CSE estimates charge states by focusing on these lines as in the case of the operation on the experimental data.
These demonstrate that the CSE estimates charge states by focusing on these lines, indicating that human-like recognition can be realized.

\begin{figure}
\begin{center}
  \includegraphics{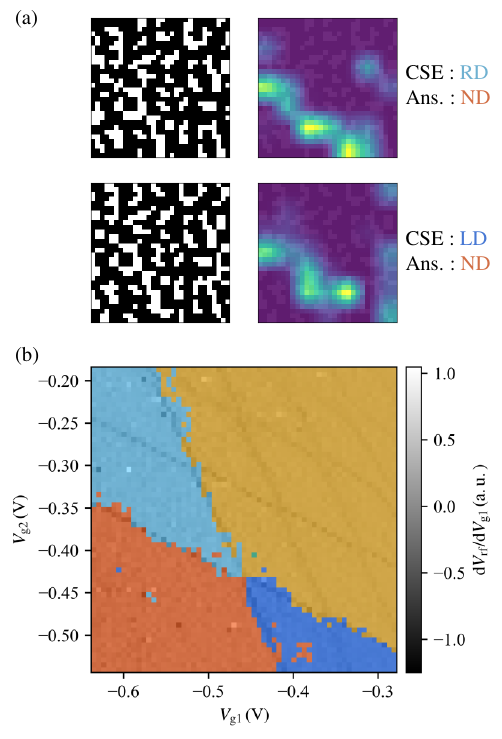}
  \caption{(a) The results of Grad-CAM for misclassifications region by the CSE. 
  (b) The charge state recognition of experimental data by improved CSE.}
  \label{fig4}
\end{center}
\end{figure}

Here, it is important to note that in Fig.~\ref{fig2}, there are several instances where regions labeled as ND are incorrectly classified as RD and LD.
Such misclassifications outside the boundary regions of charge states pose a problem in the auto-tuning of the quantum dots.
Examining the Grad-CAM results at these misclassification regions, it appears that areas where a few noise pixels are connected are identified as charge transition lines as shown in Fig.~\ref{fig4}(a).
To address this issue, we increased the ND training data by fourfold to balance the data with and without charge transition lines, further teaching the CSE that noise pixels can occasionally connect in the ND state.
As a result, as shown in Fig.~\ref{fig4}(b), the misclassifications in the ND region have notably decreased.
This demonstrates that the feedback from the Grad-CAM is useful to improve the model.
The accuracy are also evaluated as $(92.8 \pm 1.3)\%$, $(96.1 \pm 0.3)\%$, and $(85.7 \pm 1.2)\%$ for our experimental, simulation, and other group's data, respectively.
Note that these accuracy values are primarily influenced by misclassifications near the boundary regions of the charge states. 
Then modifying the training data resulted in negligible effects on the overall accuracy, even though it improved classifications outside these boundary regions.


\section{Summary}
In this paper, we analyze the explainability of the machine learning model estimating charge states in quantum dots. 
To generate the training data, we employ the CI model which is a simple simulation model and clearly shows the charge transition lines we typically focus on to recognize the charge states.
The CSE recognizes five charge states in both our experimental data and that of other group, and its interest highlighted by Grad-CAM is on the charge transition lines, the same as humans.
This allows for the tuning of CSE parameters, such as training data, based on human insights and has been empirically demonstrated to result in improvements.
This approach eliminates the need for excessive parameter tuning, thus reducing the cost associated with generating an optimal ML model.
We also thoroughly detail the pre-processing steps for both training and experimental data, offering sufficient information to enable replication of our method.

Toward large-scale semiconductor qubits, the method is needed to recognize the charge state in the additional multiple quantum dots.
One of approach is utilizing a virtual gate~\cite{Liu2022}, which can translate the charge stability diagram of the multiple quantum dots into that of the double quantum dots.
By utilizing virtual gates, the method demonstrated for double quantum dots in this study can also be applied to multiple dots, expanding its potential applicability.
Additionally, our pre-processing steps, including the CI model, facilitate the expansion of the quantum dot system without incurring significant additional simulation costs. 
This approach is also suited for future scaling up of quantum dot systems.


\section{Supplementary material}
See the supplementary material for additional details regarding the CNN model. 

\section{Acknowledgments}
We thank 
RIEC Fundamental Technology Center and the Laboratory for Nanoelectronics and Spintronics
for the technical support. 
Part of this work is supported by 
MEXT Leading Initiative for Excellent Young Researchers, 
Grants-in-Aid for Scientific Research (20H00237, 21K18592, 23KJ0200), 
Fujikura Foundation Research Grant, 
Hattori Hokokai Foundation Research Grant, 
Kondo Zaidan Research Grant, 
Tanigawa Foundation Research Grant, 
FRiD Tohoku University, 
and Cooperative Research Projects of RIEC.
Y. M. acknowledges WISE Program of AIE for financial support.

\section{AUTHOR DECLARATIONS}
\subsection{Conflict of Interest}
The authors have no conflicts to disclose.

\subsection{Author Contributions}
\textbf{Yui Muto:} Data Curation (lead); Investigation (lead); Methodology (equal); Software (lead); Visualization (lead); Funding Acquisition (equal); Writing/Original Draft (lead). 
\textbf{Takumi Nakaso:} Conceptualization (equal); Investigation (equal); Methodology (lead); Software (equal); Writing/Review \& Editing (equal).
\textbf{Motoya Shinozaki:} Data Curation (equal); Investigation (equal); Methodology (equal); Visualization (equal); Resources (lead); Writing/Original Draft (equal); Writing/Review \& Editing (equal). 
\textbf{Takumi Aizawa:} Investigation (equal); Methodology (equal); Resources (equal); Writing/Review \& Editing (equal).
\textbf{Takahito Kitada:} Investigation (equal); Methodology (equal); Resources (equal); Writing/Review \& Editing (equal).
\textbf{Takashi Nakajima:} Methodology (equal); Resources (equal); Writing/Review \& Editing (equal).
\textbf{Matthieu R. Delbecq:} Methodology (equal); Resources (equal); Writing/Review \& Editing (equal).
\textbf{Jun Yoneda:} Methodology (equal); Resources (equal); Writing/Review \& Editing (equal).
\textbf{Kenta Takeda:} Methodology (equal); Resources (equal); Writing/Review \& Editing (equal).
\textbf{Akito Noiri:} Methodology (equal); Resources (equal); Writing/Review \& Editing (equal).
\textbf{Arne Ludwig:} Methodology (equal); Resources (equal); Writing/Review \& Editing (equal).
\textbf{Andreas D. Wieck:} Methodology (equal); Resources (equal); Writing/Review \& Editing (equal).
\textbf{Seigo Tarucha:} Methodology (equal); Resources (equal); Writing/Review \& Editing (equal).
\textbf{Atsunori Kanemura:} Conceptualization (equal); Methodology (equal); Software (equal); Writing/Review \& Editing (equal).
\textbf{Motoki Shiga:} Conceptualization (equal); Methodology (equal); Software (equal); Writing/Review \& Editing (equal).
\textbf{Tomohiro Otsuka:} Conceptualization (lead); Methodology (equal); Resources (equal); Funding Acquisition (lead); Supervision (lead); Writing/Review \& Editing (lead). 

\section{DATA AVAILABILITY}
The data that support the findings of this study are available within the article and at https://en.qd.riec.tohoku.ac.jp/database.


\begin{references}
\bibitem{OpenCV}
The function of cv2.adaptiveThreshold is used. 
The parameters are set to adaptiveMethod = cv2.ADAPTIVE\_THRESH\_GAUSSIAN\_C, thresholdType = cv2.THRESH\_BINARY, blockSize = 5, and $C$ = -1.
\end{references}
\end{document}